# Realism and instrumentalism about the wave function. How should we choose?

Mauro Dorato,
Department of Philosophy, Communication and Performing Arts
Section of Philosophy, University of Rome Three
Via Ostiense 234, 00146, Rome, Italy

Federico Laudisa,
Department of Human Sciences,
University of Milan-Bicocca
Piazza Ateneo Nuovo 1, 20126, Milan, Italy

Abstract

The main claim of the paper is that one can be 'realist' (in some sense) about quantum mechanics without requiring any form of realism about the wave function. We begin by discussing various forms of realism about the wave function, namely Albert's configuration-space realism, Dürr Zanghi and Goldstein's nomological realism about Ψ, Esfeld's dispositional reading of Ψ and Pusey Barrett and Rudolph's realism about the quantum state. By discussing the articulation of these four positions, and their interrelation, we conclude that instrumentalism about Ψ is by itself not sufficient to choose one over the other interpretations of quantum mechanics, thereby confirming in a different way the indetermination of the metaphysical interpretations of quantum mechanics.

Key words: wave function, configuration space, nomological realism, quantum state realism, dispositionalism, undeterdetermination

1 Introduction

It is not exaggerated to claim that one of the major divides in the foundations of non-relativistic quantum mechanics derives from the way physicists and philosophers understand the status of the wave function. On the instrumentalist side of the camp, the wave function is regarded as a mere instrument to calculate probabilities that have been established by previous measurement outcomes.[1] On the other "realistic" camp, the wave function is regarded as a new *physical* entity or a *physical* field of some sort. While both sides agree about the existence of quantum "particles" (the so-called theoretical entities), and therefore reject the radical agnosticism about them preached by van Fraassen (1980), the various "realistic" (and consequently, instrumentalist) philosophies of quantum mechanics are typically formulated in different, logically independent ways, so that their implications need to be further investigated.

---

[1] Among representative of this form of instrumentalist, one can cite, among many others, Bohr (1972-2006) and Rovelli (1996). For Bohr's antirealism about quantum theory (and realism about quantum entities) see Faye 2001. For Rovelli's analogous stance, we refer the reader to Dorato 2013. For an exposition of Rovelli's relational interpretation, see Laudisa and Rovelli 2013. Here, we don't worry about the tenability of the distinction between entity realism and theory realism.

For instance, on the one hand it seems plausible to claim that a realistic stance about the wave function is not the only way to defend "realism" about quantum theory. One can support a "flash" or a "density-of-stuff" ontology (two variants of GRW), or an ontology of particles with well-defined positions (as in Bohmian mechanics), as *primitive* ontologies for observer-independent formulations of quantum mechanics (Allori, Goldstein Tumulka, Zanghì 2008). "Primitive ontologies", as here are understood, are not only a fundamental ground for other ontological posits, but also entail a commitment to something concretely existing in spacetime (see also Allori 2013). On the other hand, however, it is still debated whether such primitive ontologies can be autonomous from some form of realism about the wave function (Albert 1996).

In order to discuss this problem, we begin with a preliminary clarification of the meaning of "realism" and "instrumentalism" in physics, which are often subject to ideological and abstract discussions that often have little to do with the practice of physics (section 2). In the following sections we present the various forms that a realism about the wave function can take; namely, in section 3 we assess configuration-space realism (Albert 1996), or wave function-space realism (North 2013), a form of realism that might be backed up by Psillos's (2011) realist, and "literalist" attitude toward the abstract models postulated by physics. In section 4 we discuss what we call $\Psi$-nomological realism – or realism about the guiding law of Bohmian mechanics – as a consequence of a more general primitivism about physical laws defended in Maudlin (2007)[2]. Considering the wave function of the universe as a *nomological object* is a way of defending this position (Goldstein and Zanghì 2013, p.96). In section 5 we present a form of *indirect* wave-function realism, according to which the wave function indirectly refers to real physical properties, for instance in virtue of the eigenvalue-eigenvector link: "the wave function doesn't exist on its own, but it corresponds to a property possessed by the system of all the particles in the universe" (Monton 2006, p. 779). The recent dispositionalism about quantum properties seems a way to formulate this position (Dorato 2007b, Dorato and Esfeld 2010, Esfeld, Hubert, Lazarovici and Dürr 2013). In section 6 we evaluate a much debated wave-function realism, according to which the quantum state (as described by the wave function) is independent of the knowledge of the observer, so that it is more than mere "information" that observers have about the system (Pusey, Barrett, Rudolph 2012).

A natural question is which of these various ways of formulating realism about the wave function (RWF for short) is more plausible, in the hypothesis that they are all independent of each other. Providing an answer to this question (and therefore to the problem whether instrumentalism about the wave function is not the most reasonable position to take) is the main target of our paper.

---

[2] "I suggest to regard law as fundamental entities in our ontology" (Maudlin 2007, p. 18).

2. Realism as a stance and its pluralistic consequences

Is it possible to discuss the ontological status of wave function independently of a specific interpretation of quantum mechanics? In order to answer this question in the affirmative, some considerations on the realism/instrumentalism debate seem appropriate[3]. In our opinion, to be a realist about physical theories in general is a *stance* (van Fraassen 2002) that is, an attitude toward the aim of physical theorizing. This assumption entails that there is no *a priori* guarantee that such an aim will be accomplished in all cases or by all theories. Often, scientists and philosophers are able to tell – and history can teach us – when a realist approach to a given theory is justified or not. It then becomes not unreasonable to be instrumentalist about physical theory $x$ and realist about theory $y$, according to the kind of evidence (and other epistemic virtues) that $x$ and $y$ can boast.[4] Even more radically, one can be instrumentalist about *different components* of the *same* physical theory: Lange (2002), for example, argues convincingly that one ought be realist about the electromagnetic field but antirealist about Faraday's lines of force.

If we adopt the above-mentioned anti-ideological and pragmatic attitude toward scientific realism in general, an evaluation of the *pros* and *cons* of the various kinds of wave-function realism does not *a priori* force us to take a stand in favor or against a particular type of a primitive ontology for quantum theory. Our inquiry can be important for evaluating the different interpretations of quantum mechanics with respect to the status of the wave function; these interpretations and their mutual relations in fact cannot be represented exclusively in terms of logical implications between the above mentioned primitive ontologies (PO) and the different forms of wave function-realism (WFR). To exemplify, let us consider the two following possibilities:

1) Let us suppose that the assumption of a primitive ontology requires some form of realism about the wave function as a necessary condition (PO→WFR). If this were the case, instrumentalists about the wave function could reject primitive ontologies *via* a simple *modus tollens*. In this first alternative, the question of inquiring into the reality of the wave function *per se* assumes a particular importance, but Bohmian mechanics turns out to be a counterexample to the claim that we *need* to treat the wave function as a robustly real entity in its own right in order to be justified in assuming a primitive ontology.

2) The converse implication (WFR → PO) amounts to assuming that an attribution of some type of ontological status to the wave function presupposes a primitive ontology of a sort as its necessary condition. Again, there seems to be a counterexample to the complete generality of such an implication: in the configuration space realism defended by Albert, elevating the configuration space in which the wave function lives to the status of ultimate reality need not imply the requirement of a primitive ontology of entities in spacetime as primary objects the theory is about, since the theory in Albert sense is *primarily* about the configuration space itself.

---

[3] For lack of space, here they will be have to be taken for granted.
[4] For this viewpoint, see Dorato (2007a).

In both cases, anyway, establishing in what sense the wave function can be an "element of reality" will have interesting implications for the kind of primitive ontology that is more plausible to adopt. Since these brief remarks should suffice to justify the focus of our paper on wave function-realism, we can proceed to discuss the various options at stake.

## 3 Realism about configuration space

John Bell once wrote about quantum mechanics: "no one can understand this theory until he is willing to think of ψ as a real objective field rather than just a 'probability amplitude.' Even though it propagates not in 3-space but in 3N-space" (1987, 128). David Albert takes inspiration from this passage, as in his view (1996, 2013), the wave function is regarded as a physical *field*. It is often presupposed that since any physical field is an assignment of values to a space, the space on which the field sits must be regarded as real. As is well known, however, the wave function can be an assignment of physical magnitudes (positions, for Bohmian mechanics) to every point of 4d spacetime only if we have a one-particle system. As soon as *N* particles are considered, the wave function lives in a 3*N* configuration space: "The sorts of physical objects that wave functions *are*, on this way of thinking, are (plainly) *fields* - which is to say that they are the sorts of objects whose states one specifies by specifying the values of some set of numbers at every point in the space where they live, the sorts of objects whose states one specifies (in *this* case) by specifying the values of *two* numbers (one of which is usually referred to as an *amplitude,* and the other as a *phase)* at every point in the universe's so-called *configuration* space" (Albert 1996, p.277).

One widely recognized, first problem with this view, is how one can recover tables, chairs occupying a 4-dimensional spacetime (namely POs in the sense of Allori et. al 2008) from a 3*N* dimensional configuration space. Using magic words like "emergence" is not going to help: until a convincing explanatory sketch of such an emergence is available, we submit that one has no reason whatsoever to take configuration space realism seriously.

It could be replied that while science is a sophistication of common sense, it is often capable of reaching conclusions that cast radical doubts on important *components* of common sense. Our first argument against this reply is that the stress in the previous sentence is on "components". Notice the difference with past episodes in the history of science. For example, when natural philosophers discovered that the Earth is not stationary, they had to explain how it could be in motion without us noticing it. The reconciliation of the scientific worldview with the world of our senses was achieved via the introduction of the notion of inertia. An analogous explanation was achieved of our natural belief in the worldwide nature of the present moment, which was later superseded by Einstein's postulation of the relativity of simultaneity. In fact, one can explain why we tend to believe that the present moment has cosmic extension in terms of the speed of light and the finite duration needed by our brain to process temporally successive light signals (Dorato 2011).

However, in the case of configuration space realism, it is the whole worldview of common sense that is regarded as "misleading", and since science relies on observations and therefore on common sense, the consequence that all our observations are radically illusory cannot be accepted.

It must be admitted that quantum mechanics requires anyway an important sacrifice of elements of the manifest image; but in this regard even Everettian quantum mechanics is in better shape with respect to the task of explaining the emergence of our spacetime from configuration space, insofar as it can explain with the help of decoherence why the local observer cannot perceive any interference with the other worlds. In other words, if believing that the wave function is a physical entity (a field) implies configuration realism, it could be argued by *modus tollens* that the wave function is not physically real. Since the abstract or concrete ontological status of the wave function will be discussed in later sections, let us assume that the wave function might be neither concrete nor abstract, and yet a wholly new physical entity (Maudlin 2013). After all, why should we assume that something is physical only if it is in 4D spacetime?

Two remarks are sufficient to create troubles to this assumption. First, the case of strings, which live in compactified dimensions, is different from that of a $3N$ dimensional field. Strings still live in spacetime, even though the latter is conceived as being ten-dimensional or even 26-dimensional. The fact is that the extra dimensions are too small to be "seen" (compactification). The second difficulty is given by the fact that the problems that afflict configuration space realism also arise in the case of a multidimensional ($3N$) physical field. How can we recover a four-dimensional field (say the electromagnetic field) from the former? Until an explanatory sketch is provided, there is no reason to reify the wave function by requiring that the mathematical space needed to define it is the real stuff the universe is made of. As Maudlin notes (2013, p.152), mathematical representations of physical phenomena are not a clear guide to ontology, since they often do not guarantee even isomorphic relations between themselves and the latter. Furthermore, for obvious algorithmic reasons they must greatly simplify and idealize the target they are a vehicle for, and so they are not necessarily similar to what the are supposed to denote.

A different form of realism about the wave function has been defended by North (2013), who distinguishes between *configuration*-space realism and *wave*-function realism, a kind of ontic structural realism about the latter. Here we can afford to be brief about her interesting proposal, since she assumes rather than argue that the wave function "directly represents or governs" parts of the ontology of quantum mechanics (ibid., p.185). Her main stance is a form of epistemic primitivism about laws, since she claims that dynamical laws of a theory are our main guides to infer what exists according to the theory, and what exists at the fundamental level is the structure that is needed to formulate the laws. What is missing in her semi-transcendental approach is the validation of the claim that there is only *one* mathematical way to formulate the dynamic laws, a step that is necessary to claim uniqueness also for the inferred physicalstructure. In fact she denies any guiding role to Hilbert space (ibid, p.191) and she does not even mentions other more algebraic and abstract formulations of quantum mechanics; but it is not wholly clear on the basis of which criterion this selection is suggested: whether a state space has too little or superfluous structure typically depends on the problem at hand. And we want to add that not by chance such "uniqueness questions" are a typical problem for any form of ontic structural realism, since it is highly difficult to prove that the same dynamical laws cannot be formulated by presupposing a different mathematical structure.[5]

4 The wave function as a nomological entity

In Goldstein and Zanghì (2013), the wave function is defined as a "nomological entity", the primitive ontology being constituted by the positions of particles in spacetime, or by the actual positions of the particle $\mathbf{Q} = (Q_1, Q_2, Q_3, Q_4, \ldots Q_n)$ in configuration space. Since the two authors are not terribly clear about what we should mean by "nomological entity" (are physical laws entities?), it is important to defend their position as best we can in order to overcome initial resistances of philosophers to the infelicitous choice of the term.

First, evidence for a robust ontic status of $\Psi$ is suggested by its role in Bohm's "guiding" equation: the velocity of any of the *N* particles is a function via $\Psi$ of the positions of all the other particles. Second, according to Goldstein and Zanghì, the real nomological entity is properly speaking only the wave function of the universe, since the universe is " the only genuine Bohmian system" (ibid., p. 94), the wave function of a subsystem being only definable in terms of the wave function of the universe and the whole set of configuration of all the particles. Given the fundamental non-locality of the theory, this is only to be expected, even though for all practical applications what one deals with in Bohmian mechanics are subsystems. Since this presupposes the possibility of attributing a wave function to the universe, it seems legitimate to ask whether this move is legitimate[6], given the present lack of a quantum theory of gravity, or even of a well-worked out relativistic extension of Bohmian mechanics. Despite the fact that at the moment the attribution of a wave function to the universe is rather speculative, or even devoid of any clear empirical meaning, for the sake of the argument we will assume without further ado that our ontological quest is limited to a Newtonian, non-relativistic spacetime, which possesses a privileged foliation.

Given these two clarifications, the real question (our third point) is of course how to understand the ontological status of the wave function as a nomological entity. It will not do to invoke vague metaphors like the wave functions "choreographs" or "governs" the motion of the particles, since laws strictly speaking do not govern like kings: if they literally governed, they would have to be "external" (to continue the metaphor) to what they govern. But if they are external, how can they affect physical entities in the sense in which $\Psi$ must "guide" the motion of particles? This governing view seems a remnant of a theological, prescriptive rather than descriptive conception of laws, motivated by the hypothesis that a Creator imposes his own will to Nature, its creature (Dorato 2005, chapter 1).

---

[5] This problem is no less acute in spacetime theories, where general relativity can be formulated in a variety of different mathematical formalisms (that of Riemannian differentiable manifolds, Einstein algebras, twistors, non-commutative geometry and so on).
[6] In the Rovelli relational interpretation, for instance, such an attribution makes no sense (Laudisa, Rovelli 2013, Dorato 2013).

Abandoning, as it is fair to do, the literal interpretation of the term "governing", there is still an important question that needs to be raised *a propos* of the wave function regarded as a "nomological entity": are nomic entities in general external or internal to the entities and the properties that instantiate them? This issue is important in order to clarify the property-first view of laws *vs.* nomic primitivism, and therefore how we should understand Zanghì and Goldstein's view of Ψ as nomological. Moreover, it leads to specifying three different senses of 'primitive': the first refers to the primitive ontology of space-time located entities the theory is about, the second refers to the *conceptually* non-reductive character of the notion of lawhood which the primitivism about laws is grounded upon and, finally, the sense in which such 'special' nomological entities as the wave functions are *ontologically* primitive.

Non-metaphorically, the term "external", when referred to laws, typically means "independent or non-supervenient upon the entities and the properties they relate", while "internal" is therefore equivalent to "dependent on those entities and properties". As Psillos put it, external means that the laws can vary while the properties that they instantiate do not change (2006, p. 18) and this implies a sort of *quidditism*. This is the view that there are properties $P$ whose identity is independent of, and can be detached from, their nomic or causal role $R$, so that it is not essential to a property that it plays a given nomic role. It should be admitted that quidditism, exactly as heacceitism, cannot be ruled out *a priori*. However, it is certainly difficult to accept the view that the property $P$ that electrons possess of being negatively charged – which entails the nomic role $R$ to attract positively charged bodies – could be detached from $R$ in such a way that $P$ would remain the *same* even if governed by a different law (and therefore be characterized by a different $R$). Be that as it may, the other horn of the dilemma (laws as internal to properties) implies "a property-first" view on laws, and therefore the idea that laws supervene on properties and relations of entities and cannot be ontically primitive, let alone "govern" their instances.

Leaving aside the metaphysical complications of quidditism, for us it is important to note that the choice between these two alternatives ("externalism" or "internalism" about laws) does not force one to be antirealist about laws,[7] a position that would rule out the possibility that Ψ be a nomological entity in the sense of Goldstein and Zanghì. Nevertheless, in the remainder of this section we will assume that their view is committed to primitivism (or the non-supervenience of laws on properties) for essentially two reasons. First, the "internal", second alternative pushes toward nomic antirealism, since the properties or the powers of entities could exhaust all the roles played by laws (Mumford 2004). Second, the property-first view of laws implied by internalism will be discussed in the next section.

---

[7] One could claim that laws exist but that they are just relations between entities, that are primary and more fundamental.

Suppose then that wave function realism is committed to some sort of *ontic* primitivism about laws in the sense of Maudlin (2007). The problem is that once one abandons the safe ground offered by the *conceptual* priority (or irreducibility) of nomic concepts in the sense of Carroll (1994) (the second sense of primitive above), it is not clear what *ontic* primitivism amounts to. On the one hand, we cannot assume without further arguments that conceptual priority entails ontic priority, since the concept of *scientific law* might be irreducible to other related concepts (causation, counterfactuals, regularity, etc.), without implying any sort of ontic primitivism about laws of nature.[8] On the other hand, if one does not want to beg the question against primitivism, it must be admitted that there is a sense in which ontic nomic primitivism cannot be further understood, precisely because the notion of law is regarded as un-analysable.

However this irreducibility might be regarded as a serious deficiency of this position for at least two reasons.

1) It is true that we must start from somewhere, in mathematics as well as in philosophy: it is the explanatory consequence of taking a notion *A* as primitive that justifies the choice of *A*. However, mathematics relies on axioms, which give an implicit definition of the axiomatized notions. In philosophy, on the contrary, when we do not understand a notion (in this case 'laws of nature regarded as existent'), we seem to be in a different and more difficult predicament. When a concept *A* is more obscure than a concept *B*, and we declare *A* "primitive" – laws seem to be less intuitively understood than properties – we run the risk of wanting to solve a philosophical problem without even trying.

This difficulty, however, can be solved: after all, intuitions about what is obscure may vary. Let us then agree that a fair reading of "ontically primitive" with respect to laws might mean, simply, that *there are mind-independent nomic facts* that are the supervenience basis for the existence of those properties, dispositions, causal facts and the like that (according to the primitivists) are mistakenly regarded as the truth-makers of the propositions that express the "laws of science".

2) This formulation brings with itself the second difficulty. Since these (approximately true) propositions regarded as truth-bearers in physics are typically differential equations, for the primitivist about laws the existence of *nomic* (physically necessary) facts must be contrasted with the existence of merely *contingent* facts, typically lying in hypersurfaces of simultaneity, and specifying the initial or boundary conditions to which the equations are applied. But how can the primitivist distinguish between the modally loaded, nomic facts, and the contingent facts, if both are *facts*? Clearly, ontic primitivists about laws cannot ground the distinction between nomic and contingent truths on the existence of physically possible worlds, lest law loses its primitivity. Same for purely conceptual primitivism. Furthermore, note that in this rendering, ontic primitivism has to be realist about the existence of facts and must regard them as *concrete* entities, being in any case distinct from *abstract* propositions.

---

[8] The concept of knowledge might be irreducible to justified true belief, and yet knowledge is not ontically primitive.

If we apply these two objections to our problem, the difficulty should be obvious: claiming that the wave function is a entity because the laws in which it appears exist in a primitive sense is not convincing, because a physical hypothesis is made to depend on a highly controversial metaphysical hypothesis.

5 The property-first view of the wave function: dispositionalism

We have seen that according to *primitivism* about, say, the guiding law of Bohmian mechanics, there is in the quantum world a global, nomic fact instantiated by the world in question that determines the temporal development of an initial, contingent configuration of particles belonging to an hypersurface of simultaneity. According to the property-first view, it is instead the initial configuration of point-particles in a background spacetime that, by instantiating a plurality of properties, fixes the temporal evolution of whatever exists in the initial configuration. In the literature, such properties have often been regarded as dispositions, so that the subsequent behaviour of the initial configuration of particles is given by their manifestations.

Just to exemplify, in the case of flashy GRW, the disposition of non-massless entities to localize in a flash, or in a region of spacetime in the mass-density reading of GRW, is a spontaneous and an irreversible process. The flash or a certain localized mass density in spacetime is the manifestation of the disposition in question. In the case of Bohmian mechanics, each particle has a spontaneous disposition to influence the velocity of the i-th particle in a non-local way, and the velocity of that particle is the manifestation of the global disposition carried by the whole configuration of particles (Dürr, Esfeld, Lazarovici, et. al. 2013). Thus, on Bohmain mechanics the configuration of *all* particles at a given time *t* instantiates a *dispositional property* that manifests itself in the velocity of each particle at *t*; the universal wave-function at *t* represents that property, so that the latter is ontologically primary and the wave function refers to such a property.[9]

The difference with GRW's two primitive ontology is that the dispositions in the latter are really probabilistic propensities (GRW is irreducibly indeterministic), while in Bohmian mechanics they are sure-fire dispositions. But in both cases (deterministic and indeterministic), the introduction of dispositional properties has the advantage of avoiding a reification of the configuration space.

However, in both cases there are two difficulties that all these property-first views share: i) quantum dispositions are spontaneous but in standard situations classical, typical dispositions need a stimulus (a stone breaking a window pane, with the ensuing manifestation of fragility being the broken

---

[9] For a more detailed description of this view, see Dorato and Esfeld, forthcoming

glass); ii), physical laws, referring to or representing dispositions, are, unlike dispositions, *time-symmetric*.

The first difficulty (i) depends on how one defines dispositions, namely in a more liberal or in a less liberal way, so as to encompass also spontaneous manifestations. We think liberalism about this issue can be justified, in order not to gen the question against dispositionalism. Any mass has a spontaneous disposition to move inertially, even though the disposition to resist acceleration is manifested only in the presence of a force (the stimulus). Likewise a radioactive material has a spontaneous disposition (a propensity) to decay, even though the decay can be accelerated by bombarding the nucleus in a opportune way.

In the two GRW cases, the second difficulty (ii) could be more easily accommodated by treating the new non-linear equations introduced by the dynamical reduction models as time-asymmetric *laws*, namely nomic irreversibilities that explain or ground the less fundamental arrows of time (as suggested by Albert 2000). In the Bohmian case, the two main laws are time-symmetric, but one can hold that the irreversible dispositions to influence the velocity of each particle correlates to the arrow of becoming, the successive occurrence of events given by the manifestation of the dispositions. In this way, only one of the two directions of time is the one in which the world unfolds, so that the temporal symmetrical feature would involve only the laws of *science* and not the laws of *nature*, which would take part in a universal process of becoming. Such a process can be regarded as either primitive (Maudlin 2007, Ch. 3), or explained by the manifestation of the various dispositions making true the laws of science (for the distinction between laws of science and laws of nature, see Weinert 1995).[10]

6. The PBR theorem

As far as the controversy over the nature of the wave function is concerned, a new twist to the debate was provided by the so-called PBR theorem (Pusey, Barrett and Rudolph 2012). According to a natural reading of this result, assuming the wave function of a quantum system *S* as a mere catalogue of the information available about *S* implies predictions that contradict those of quantum mechanics. As a consequence – we might argue – the idea that a quantum state is not just information about an entity but is a real entity *itself* should be taken seriously on physical and mathematical grounds. As a matter of fact, neither the general framework of the theorem nor the specific assumptions under which it is proved are innocuous, but before attempting an assessment let us recapitulate the result. The main hypothesis on the background is that "a system has a 'real physical state' – not necessarily completely described by quantum theory, but objective and independent of the observer" (Pusey, Barrett and Rudolph 2012, p. 475). That such a state might be not completely characterized by quantum theory implies that a wave function ψ for a system *S* is taken to represent a *preparation* of the system itself: ψ fixes the 'real' state λ non-uniquely but rather according to a probability distribution $\mu_\psi(\lambda)$. In the PBR approach – inspired by the terminology introduced in Harrigan, Spekkens (2010) – given two wave functions ψ and ϕ, we have two possible alternatives: either the probability distribution $\mu_\psi(\lambda)$ and $\mu_\phi(\lambda)$ do overlap or they do not. In the former case, there are values of the distributions that might be assigned to both ψ and ϕ, something that testifies an uncertainty on what the 'real' state associated either with ψ or ϕ might be; in the latter case, the non-overlapping testifies the lack of uncertainty: "informally, every detail of the quantum state is 'written into' the real physical state of affairs" (PBR 2012, p. 476: Harrigan, Spekkens 2010 speak of *epistemic* view in the former case and of *ontic* view in the latter). Under the additional assumption that independently prepared systems have independent physical states[11], PBR prove that, given two distinct quantum states ψ and ϕ, the overlapping of the respective $\mu_\psi(\lambda)$ and $\mu_\phi(\lambda)$ implies a contradiction with the statistical predictions of quantum mechanics.

---

[10] This point has been initially suggested in Dorato and Esfeld (forthcoming)

The PBR theorem is the n-th result of a long chain of *no-go theorems*, namely results that in principle should clarify the fundamental structure of the theory, by pointing out the boundaries that the theory itself is supposed not to violate when satisfying a class of basic constraints. Even leaving aside the general significance of the no-go strategy in the foundations of physics (Laudisa 2014), there are several critical points that need be emphasized. The first is the most obvious but, nevertheless, the most urgent: one may ask what is the meaning of the assumption according to which 'a system has a real physical state' when we lack a clear understanding of what it *means both* for the wave function ψ and the 'real' state λ "to be real"[12]. If it means that it is more than mere information, we still haven't been told much, that is, we have not been told what is it and what its properties are. As a consequence, the lack of a clear notion of what it takes for states like ψ or λ to be 'real' implies that it is also completely unclear what it means that we *cannot* interpret ψ as mere information. It might be argued that, when we have ontic models, the quantum states *supervene* on 'real' states, namely no change in quantum states without change in real states. Does this 'supervenience' talk, however, help in understanding what it means to be 'real', in absence of an ontologically clear formulation of quantum mechanics itself? In some sense, both ψ– and λ–sort of states are supposed to carry with themselves an ontological stock that they in fact are unable to justify. For consider even the case of classical mechanics, that in the PBR approach is taken into account in order to explain the epistemic-versus-ontic view of states. In the Newtonian dynamics of a single point particle in one dimension, the description consists in specifying a point in the relevant phase space, namely a pair $\langle x(t_0), p(t_0)\rangle$ at some given initial time $t_0$, where $x(t_0)$ is a position value and $p(t_0)$ is a momentum value at $t_0$: under the ideal assumption that we know all the forces at work, we can determine any pair $\langle x(t), p(t)\rangle$ at any given time $t$ by using either the Newtonian or the Hamiltonian formulation of the dynamical laws. Now, it seems very natural at first sight to make sense of a pair like $\langle x(t), p(t)\rangle$ by stating that it is a clear instance of an 'ontic state', namely of 'a state of reality'. Since, however, that pair is in fact simply a point in an abstract, multidimensional configuration space, it must be noted that in order for such a 'classical' pair to be a 'state of reality', a rather heavy assumption must be accepted, namely that what is 'really' real is not our three-dimensional experience but rather the manifold with an astronomical number of dimensions whose points are all the possible ontic states determined by the classical dynamical laws. Therefore this reference to the supposedly more familiar case of classical mechanics on phase space, far from serving the purpose of PBR of enlightening the meaning of what a 'real' state should be, shows that there is a big gap between certain mathematical structures of the state space on one side and the realm of 'real' states (whatever they might be).

---

[11] That this assumption is indeed necessary is proved in Lewis, Jennings, Barrett, Rudolph 2012; see also Schlosshauer, Fine 2012.

[12] That this is a problem can be seen also if we realize that (as PBR themselves remark) an instrumentalist is allowed to ignore the result of the PBR theorem, unlike the case, for instance, of the measurement problem, which is at least partly a problem *also* for the instrumentalist.

7. Conclusions

After presenting and discussing the main options available on the status of the wave function in the foundations of quantum mechanics, it may be worthwhile trying to recapitulate the general framework and possibly to draw some connections among the different approaches. As far as the latter are concerned, they display all a significant metaphysical flavor inasmuch as they all adopt robust metaphysical assumptions concerning their respective target entities – namely the configuration space, the laws of nature and natural properties: as to their respective plausibility, they can be evaluated on how well they fare with respect to our intuition and common sense on one side and to the role they might play in the foundations quantum physics on the other.

If we start with the configuration space realism, we might argue that the weight of the usual objection – according to which the reality at the level of the configuration space would be hard to reconcile with the reality at the level of our ordinary, three-dimensional experience of physical systems and processes – is debatable. While common sense certainly sides with three-dimensional experience, it is also true that the extent to which commonsense is really 'common' might be controversial. Since commonsense is a vague notion, certain tenets of common sense might be subjective, as different people might have different views on what counts as important in 'commonsense'. Furthermore, the abstractness of the configuration space might also not be an unsurmountable problem in itself (one can adopt, for instance, the motivations defended by Psillos 2011).

The above-mentioned remark by Maudlin, however, (namely that mathematics is not often a safe guide to ontology) should be taken into due account (examples abound: recall either the algebraic formulation of quantum field theory or the quantum logic program). Moreover, a serious problem with the configuration space realism can be the attempt of supplementing the thesis that the only reality is configuration space with the highly controversial assumption of Humean supervenience (HS for short; see Loewer 1996, Darby 2012): the latter assumption would be supposed to dispense with the problem of what is the sense in which a wave function 'lives' and plays its statistical and dynamical role in a configuration space, since according to a configuration-space+HS stance only the Lewisian mosaic of facts would exist and nothing else (putting aside the problem of what is a 'fact' in configuration space: a point $x = (q_1, q_2,....)$?, the fact that a point $x$ has coordinates $(q_1, q_2,....)$? or what?)

That we need not worry if, for good reasons, we are led to include abstract entities in the inventory of the world seems to apply also to the nomological approach to the wave function. If we are prepared to contemplate a 3N-dimensional manifold as the ultimate physical reality, no less prepared should we be to contemplate the existence of nomological 'entities', whatever they might be. But this puts Goldstein and Zanghì nomological entity-realism on the same footing as Albert configuration space realism. Both bet on abstract entities. Moreover on the face of possible objections – some of which have been mentioned before – to the primitivist view of laws, on which the nomological approach seems to be most naturally grounded, it might be remarked anyway that the primitivist view seems to have at least an advantage over the property-first view in terms of both conceptual and metaphysical economy. In fact, the dispositional reading of the property-first view applied to quantum mechanics seems to imply a worldview in which the quantity of dispositional properties amounts to the quantity of particles that are supposed to display a certain behavior under certain conditions: instead of giving up laws and having an astronomical number of particles, each with its bundle of dispositional properties, would it not be 'easier' to have a restricted number of laws that account for the seemingly dispositional behavior of the particles. If, on the other, staying closer to the spirit of Esfeld at al. 2013, the many dispositions of the single particle in question really amounted to a unique disposition of the whole configuration space described by the wave function, then primitivism and global dispositionalism would seem to converge and the distinction between properties first and laws first might be purely verbal and lose some of its importance: The global nomic fact that, according to primitivism, is instantiated by the quantum world would correspond to the global disposition characterizing the configuration space in the sense of Esfeld et. al.[13]

Of course, any nominalistic philosopher of human sympathies would be inclined to reject any form of commitment to abstract entities like laws or configurations spaces, and embrace wave function instrumentalism *sic et simpliciter*. And also this latter position is certainly not incompatible with what we know about the physics of quantum theory. In a word, the most plausible moral to be drawn at this point is that the metaphysics of Ψ is radically undetermined by quantum physics and even by the sort of primitive ontology one adopts, a conclusion which need not hold for all metaphysical claims in their relation with physical theories.

---

[13] For this claim, see Dorato and Esfled (forthcoming).